\documentclass[aps,prl,twocolumn,groupedaddress,showpacs,superscriptaddress]{revtex4}
\usepackage{graphicx}

\begin{document}

\title{Feshbach-Einstein condensates}

\author{V.G.~Rousseau}
\author{P.J.H.~Denteneer}
\affiliation{Instituut-Lorentz, LION, Universiteit Leiden, Postbus 9504, 2300 RA Leiden, The Netherlands}

\begin{abstract}
We investigate the phase diagram of a two-species Bose-Hubbard model describing atoms and molecules on a lattice, interacting via a Feshbach resonance. We identify a region where
the system exhibits an exotic super-Mott phase and regions with phases characterized by atomic and/or molecular condensates. Our approach is based on a recently developed
exact quantum Monte Carlo algorithm: the Stochastic Green Function algorithm with tunable directionality. We confirm some of the
results predicted by mean-field studies, but we also find disagreement with these studies. In particular, we find
a phase with an atomic but no molecular condensate, which is missing in all mean-field phase diagrams.
\end{abstract}

\maketitle
More than eighty years ago, Einstein predicted a remarkable phenomenon to occur in a gas of identical atoms interacting weakly at low temperature and high density \cite{Einstein}. Under such conditons, when the de Broglie wavelength of the atoms becomes larger than the interatomic distance,
a macroscopic fraction of the atoms accumulates in the lowest energy state. This phenomenon, known as Bose-Einstein condensation, remained in the archives for
a long time, and was reconsidered later with the discovery of the superfluidity of Helium in 1937. It is only in 1995 with the advent of laser
cooling techniques that the first Bose-Einstein condensates of atoms were achieved \cite{CornellWieman,Ketterle}.
At present, experiments trying to achieve ultracold  and degenerate molecular gases are creating considerable excitement 
\cite{Lang,Deiglmayr,Ospelkaus}. These experiments should lead to the creation of long-lived molecular {\em Feshbach-Einstein
condensates}, with applications in the fields of precision measurements and quantum information \cite{DeMille}.

Near a Feshbach resonance, molecules are formed from atoms by tuning a magnetic field and bring into resonance
scattering states of atoms with molecular bound states \cite{Kohler}. 
In this way, conversions between atoms and diatomic molecules are induced.
A model Hamiltonian that describes mixtures of atoms and molecules was introduced and studied before
\cite{Timmermans,Dickerscheid,Dupuis,RousseauDenteneer}.
In this letter, we study this model and analyse the presence or not of atomic and/or molecular condensates, using the
recently developed Stochastic Green Function (SGF) algorithm \cite{SGF} with tunable directionality \cite{DirectedSGF}. With this exact quantum Monte Carlo (QMC) algorithm,
momentum distribution functions, which are the main indicators of condensation, are easily accessible and allow direct comparisons with experiments. We critically compare our results with the predictions of
mean-field (MF) studies \cite{Dickerscheid,Dupuis}.

We consider the model for {\em bosonic} atoms and molecules on a lattice.
The particles can hop onto neighboring sites, and their interactions are described by intra-species and inter-species onsite potentials. 
An additional conversion
term allows two atoms to turn into a molecule, and vice versa. This leads us to consider the Hamiltonian $\hat\mathcal H=\hat T+\hat P+\hat C$ with
\begin{equation}
  \label{Kinetic}    \hat T=-t_a\sum_{\big\langle i,j\big\rangle}\big(a_i^\dagger a_j^{\phantom\dagger}+H.c.\big)-t_m\sum_{\big\langle i,j\big\rangle}\big(m_i^\dagger m_j^{\phantom\dagger}+H.c.\big),
\end{equation}
\begin{eqnarray}
  \nonumber         \hat P &=& U_{aa}\sum_i \hat n_i^a\big(\hat n_i^a-1\big) \\
  \nonumber                &+& U_{mm}\sum_i \hat n_i^m\big(\hat n_i^m-1\big) \\
  \label{Potential}        &+& U_{am}\sum_i \hat n_i^a \hat n_i^m+D\sum_i \hat n_i^m,
\end{eqnarray}
\begin{equation}
  \label{Conversion} \hat C=g\sum_i\big(m_i^\dagger a_i^{\phantom\dagger} a_i^{\phantom\dagger}+a_i^\dagger a_i^\dagger m_i^{\phantom\dagger}\big).
\end{equation}

The $\hat T$, $\hat P$, and $\hat C$ operators correspond respectively to the kinetic, potential, and conversion energies. The $a_i^\dagger$ and $a_i^{\phantom\dagger}$ operators ($m_i^\dagger$ and $m_i^{\phantom\dagger}$) are the creation and annihilation operators of atoms (molecules)
on site $i$, and $\hat n_i^a=a_i^\dagger a_i^{\phantom\dagger}$ ($\hat n_i^m=m_i^\dagger m_i^{\phantom\dagger})$ counts the number of atoms (molecules) on site $i$. Those
operators satisfy the usual bosonic commutation rules.
The sums $\big\langle i,j\big\rangle$ run over pairs of nearest-neighbor sites $i$ and $j$. We restrict our study to one dimension (1D) and we choose the atomic
hopping parameter $t_a=1$ in order to set the energy scale, and the molecular hopping parameter $t_m=1/2$ \cite{RousseauDenteneer}. The parameter $D$ corresponds
to the so-called \textit{detuning} in Feshbach resonance physics.
In this paper, we will systematically use the same value $U$ for the onsite repulsion parameters and the conversion parameter,
\begin{equation}
  U=U_{aa}=U_{mm}=U_{am}=g,
\end{equation}
in order to simplify our study. It is important to note that the Hamitonian does not conserve the number of atoms
$N_a=\sum_i a_i^\dagger a_i^{\phantom\dagger}$, nor the number of molecules $N_m=\sum_i m_i^\dagger m_i^{\phantom\dagger}$, because of the
conversion term (\ref{Conversion}). However we can define the \textit{total number of particles}, $N=N_a+2N_m$, which is conserved.

While our Hamiltonian is highly non-trivial, it has become possible to simulate it exactly by making use of the SGF algorithm \cite{SGF}.
In this algorithm a \textit{Green operator} is considered,
\vspace{-0.25cm}
\begin{equation}
  \label{GreenOperator} \hat\mathcal G=\sum_{p=0}^{+\infty}\sum_{q=0}^{+\infty}g_{pq}\sum_{\big\lbrace i_p|j_q\big\rbrace}\prod_{k=1}^p\hat\mathcal A_{i_k}^\dagger \prod_{l=1}^q\hat\mathcal A_{j_l},
\end{equation}
where $g_{pq}$ is an optimization matrix, $\hat\mathcal A^\dagger$ and $\hat\mathcal A$ are \textit{normalized} creation and annihilation operators, defined as the operators that create
and destroy particles without changing the norm of the state they are applied to, and $\big\lbrace i_p|j_q\big\rbrace$ represents two subsets of site indices in
which all $i_p$ are different from all $j_p$. The Green operator is used to sample an \textit{extended} partition function,
\vspace{-0.25cm}
\begin{equation}
  \label{ExtendedPartitionFunction} Z(\beta,\tau)=\textrm{Tr }e^{-(\beta-\tau)\hat\mathcal H}\hat\mathcal G e^{-\tau\hat\mathcal H},
\end{equation}
by propagating across the operator string obtained by expanding the exponentials of expression (\ref{ExtendedPartitionFunction}) in the interaction picture.
When configurations in which $\hat\mathcal G$ acts as an identity operator occur, then (\ref{ExtendedPartitionFunction}) reduces to the partition function
$Z(\beta)=\textrm{Tr }e^{-\beta\hat\mathcal H}$, and measurements of physical quantities can be performed.
In addition, the directionality of the propagation of the Green operator is tunable \cite{DirectedSGF}, which improves considerably the efficiency of the algorithm.

An important property
of the SGF algorithm is that it works in the canonical ensemble, the canonical constraint being imposed on the total number of particles $N$.
This is essential for the efficiency of the simulations. Indeed, our model describes a mixture of two different species of particles, and would require two different
chemical potentials for a description in the grand-canonical ensemble. Adjusting numerically two chemical potentials is cumbersome, because the number of particles
of each species depends on all parameters of the Hamiltonian. Working in the canonical ensemble allows to set the total
number of particles, and the ratio between the number of atoms and molecules is controlled via the detuning, mimicking what is done in experiments.

\begin{figure}[h]
  \centerline{\includegraphics[width=0.45\textwidth]{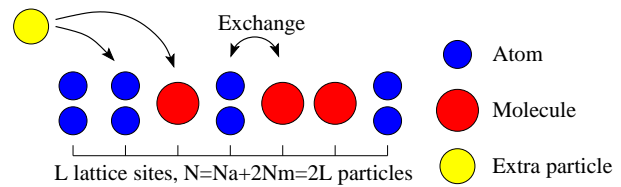}}
  \caption
    {
      (Color online) Typical configuration in a SM phase. The addition of an extra particle (atom or molecule) has a finite energy cost because it creates either
      a triplet of atoms, or an atom-atom-molecule triplet, or a pair of molecules, or an atom-molecule pair. Thus the phase is incompressible. However exchanging a pair of atoms with a molecule
      is free, thus allowing anti-correlated supercurrents. This correspond to third order processes, so the associated superfluid signals are small.
    }
  \label{Super-Mott}
\end{figure}

In order to characterize the different phases encountered, it is useful to consider the superfluid density $\rho_s$. The SGF algorithm samples the winding number $W$, so
the superfluid density is simply given by $\rho_s=\big\langle W^2\big\rangle L/2\beta$. It turns out to be more efficient to measure
$\rho_s$ by using an improved estimator $W_{\rm ext}$ \cite{SGF} for the winding number,
\begin{equation}
  W_{\rm ext}^2=\frac{2\big|\tilde j(\omega_1)\big|^2-\big|\tilde j(\omega_2)\big|^2}{L^2},\quad \omega_1=\frac{2\pi}{\beta},\quad \omega_2=\frac{4\pi}{\beta},
\end{equation}
\vspace{-0.5cm}
with
\vspace{-0.5cm}
\begin{equation}
  \tilde j(\omega)=\sum_{k=1}^n \mathcal D(\tau_k)e^{-i\omega\tau_k},
\end{equation}
where $\tau_k$ are the imaginary time indices of the interactions appearing when expanding the partition \mbox{function (\ref{ExtendedPartitionFunction})}, and $\mathcal D(\tau_k)$ equals $1$ ($-1$) if a particle jumps to the right (left)
at imaginary time $\tau_k$. This improved estimator converges faster to the zero temperature value of the winding number \cite{SGF}. In our case
we evaluate the atomic and molecular winding numbers, $W_a$ and $W_m$, and the corresponding superfluid densities are
given by $\rho_s^a=\big\langle W_a^2\big\rangle L/2\beta$ and $\rho_s^m=\big\langle W_m^2\big\rangle L/2\beta$.
In addition, it is useful to define the correlated superfluid density $\rho_s^{\rm cor}$ \cite{RousseauDenteneer},
\vspace{-0.5cm}
\begin{equation}
  \rho_s^{\rm cor}=\frac{\big\langle\big(W_a+2W_m\big)^2\big\rangle L}{2\beta}.
\end{equation}
These quantities allow to identify superfluid (SF) and \textit{super-Mott} (SM) \cite{RousseauDenteneer} phases. SF phases are characterized by non-zero values for $\rho_s^a$, $\rho_s^m$, and
$\rho_s^{\rm cor}$, while a vanishing value for $\rho_s^{\rm cor}$ with non-zero values for $\rho_s^a$ and $\rho_s^m$ is the signature of a SM phase (see caption of Fig.\ref{Super-Mott}).
\vspace{-0.25cm}
\vspace{-0.25cm}
The SGF algorithm allows to measure the atomic and molecular Green functions $\big\langle a_i^\dagger a_j^{\phantom\dagger}\big\rangle$ and $\big\langle m_i^\dagger m_j^{\phantom\dagger}\big\rangle$,
from which the associated momentum distribution functions $n_a(k)$ and $n_m(k)$ are computed by performing a Fourier transformation:
\vspace{-0.25cm}
\begin{eqnarray}
  && n_a(k)=\frac{1}{L}\sum_{pq}\Big\langle a_p^\dagger a_q^{\phantom\dagger}\Big\rangle e^{-ik(p-q)} \\
  && n_m(k)=\frac{1}{L}\sum_{pq}\Big\langle m_p^\dagger m_q^{\phantom\dagger}\Big\rangle e^{-ik(p-q)}
\end{eqnarray}

Because we are considering 1D systems, we can expect at most \textit{quasi-condensates}. These are characterized by a diverging occupation of the
zero momentum state $n(k=0)$ as a function of the size $L$ of the system, while the condensate fraction $n(0)/N$ vanishes in the thermodynamic limit. As a result,
knowing the value of the condensate fraction for an arbitrary large system size is not sufficient to determine if the system is quasi-condensed or not.
One needs to perform a finite-size scaling analysis in order to determine if $n(0)$ diverges or not.
In the following, all denoted ``condensate" phases are to be understood as ``quasi-condensate" phases. As a result, the quantities $n_a(k)$ and $n_m(k)$ allow us to
identify phases with atomic condensate (AC), molecular condensate (MC), or atomic+molecular condensates (AC+MC). It is also useful to keep in mind that the areas
below the curves $n_a(k)$ and $n_m(k)$ are exactly equal to $N_a$ and $N_m$ respectively, thus allowing an evaluation of the population of atoms and molecules.

\begin{figure}
  \centerline{\includegraphics[width=0.45\textwidth]{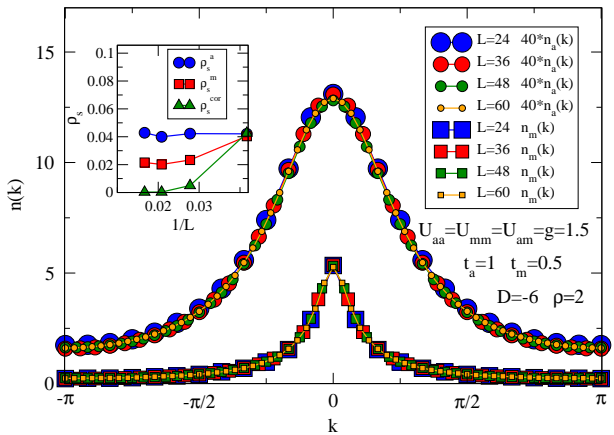}}
  \caption
    {
      (Color online) Identification of the SM phase. The occupation of the zero-momentum states
      $n_a(0)$ and $n_m(0)$ do not change as the size $L$ of the system increases, thus there is neither an atomic nor a molecular condensate.
      However the inset shows that the correlated superfluid density $\rho_s^{\rm cor}$ extrapolates to zero as the size increases, while the
      atomic and molecular superfluid densities $\rho_s^a$ and $\rho_s^m$ converge to a finite value. This is the signature of a SM phase \cite{RousseauDenteneer}.
      The errorbars are smaller than the $L=60$ symbol size.
    }
  \label{SM-Phase}
\end{figure}

We concentrate our study on systems with a total density $\rho_{\rm tot}=N/L=2$, which is one of the cases considered in MF \cite{Dupuis} and
QMC \cite{RousseauDenteneer} studies. We investigate the phase diagram in the $(1/U,D)$ plane. For sufficiently large interactions $U$, depending on the detuning $D$, we find an insulating phase characterized
by a vanishing compressibility and the absence of global superflow. This is in agreement with MF studies. However, Ref.\cite{Dupuis} denotes
this insulating phase as a regular Mott insulator (MI), whereas we find that it is actually a more exotic SM phase (see above). This can be seen in Fig.\ref{SM-Phase} for the case
$U=1.5$ and $D=-6$. The momentum distribution functions for atoms and molecules are plotted for different sizes of the lattice. No divergence of the occupation of
the zero momentum state is perceptible, so there is neither an atomic nor a molecular condensate. Moreover, the inset shows that the correlated superfluid density $\rho_s^{\rm cor}$
vanishes as the system size increases, as expected for an insulating phase. However we can see that the superfluid densities associated with the individual atomic and molecular species converge
to a finite value, which is the signature of a SM phase (a similar phase is also present in the case of Bose-Fermi mixtures \cite{Zujev}). This is the first qualitative
difference between MF results and ours.

\begin{figure}
  \centerline{\includegraphics[width=0.45\textwidth]{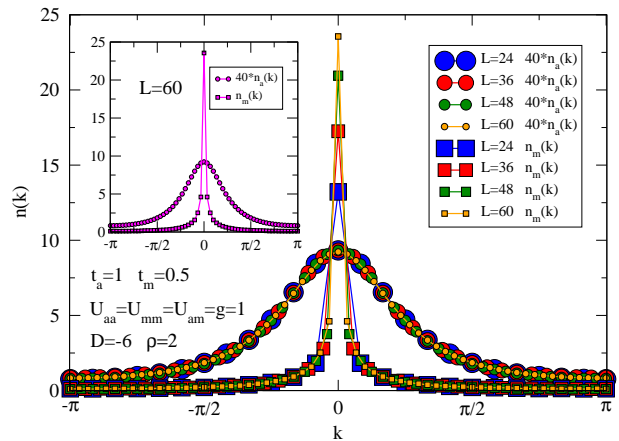}}
  \caption
    {
      (Color online) Identification of the MC phase. The occupation of the zero-momentum state of atoms
      $n_a(0)$ does not change as the size $L$ of the system increases, thus there is no atomic condensate. However the molecular momentum distribution function
      becomes narrow and $n_m(0)$ diverges as the size of the system increases, which is the signature of a MC phase. The inset shows the $L=60$ results separately
      for clarity. The errorbars are smaller than the $L=60$
      symbol size, except in $k=0$ for molecules for which the error is about 2 times the size of the symbol.
    }
  \label{MC-Phase}
\end{figure}

Starting from the above SM phase, reducing the interactions will eventually break the solid structure. For $U=1$ and negative detuning $D=-6$ the system undergoes a molecular
condensation, as can be seen in Fig.\ref{MC-Phase}. No divergence of the occupation of the zero momentum state $n_a(0)$ occurs. However $n_m(0)$
diverges and shows the presence of a molecular condensate. This transition from an insulator to a MC phase is in agreement with MF theory.

\begin{figure}
  \centerline{\includegraphics[width=0.45\textwidth]{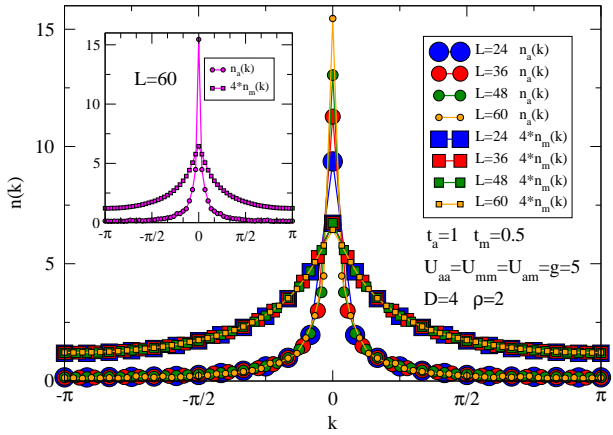}}
  \caption
    {
      (Color online) Identification of the AC phase. The occupation of the zero-momentum state of molecules
      $n_m(0)$ does not change as the size $L$ of the system increases, thus there is no molecular condensate. However the atomic momentum distribution function
      becomes narrow and $n_a(0)$ diverges as the size of the system increases, which is the signature of an AC phase. The inset shows the $L=60$ results separately
      for clarity. The errorbars are smaller than the $L=60$
      symbol size, except in $k=0$ for atoms for which the error is about 4 times the size of the symbol.
    }
  \label{AC-Phase}
\end{figure}

The SM phase persists when going from negative to positive detuning with large interactions. For sufficiently large detuning $D$, MF studies predict a
\textit{direct} transition from a MI (actually SM) phase to an AC+MC phase, as the interactions are reduced. However, our remarkable result is that we
find an intermediate AC phase in a small region of the phase diagram. This can be seen in Fig.\ref{AC-Phase} for the case $U=5$ and $D=4$. We can see that a
divergence of $n_a(0)$ occurs while $n_m(0)$ remains constant as the system size increases. Thus we are in presence of a phase in which the atoms are condensed, but not
the molecules. Such an AC phase is missing, to our knowledge, in all phase diagrams coming from MF theory \cite{Dupuis,Dickerscheid}. Our evidence for the AC phase
in Fig. \ref{AC-Phase} is comparable to that for the MC phase in Fig. \ref{MC-Phase}.

\begin{figure}[b]
  \centerline{\includegraphics[width=0.45\textwidth]{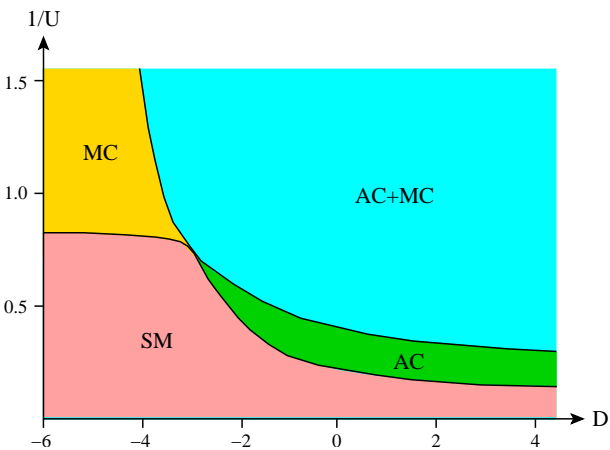}}
  \caption
    {
      (Color online) The qualitative phase diagram in the $(1/U,D)$ plane for $\rho_{\rm tot}=2$. We identify regions with super-Mott (SM), atomic condensate (AC), molecular condensate (MC),
      and atomic+molecular condensate (AC+MC) phases.
    }
  \label{Phase-diagram}
\end{figure}

\begin{figure}[t]
  \centerline{\includegraphics[width=0.45\textwidth]{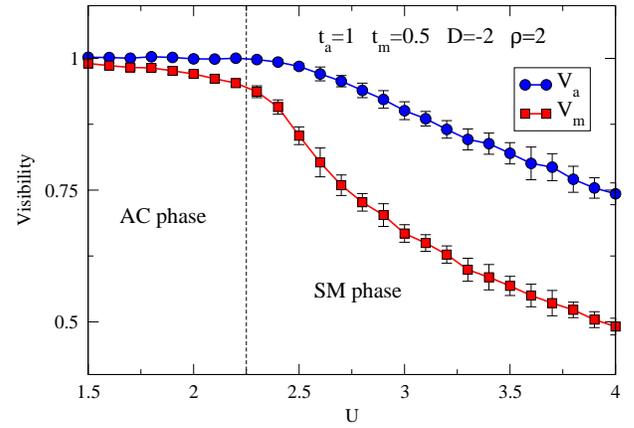}}
  \caption
    {
      (Color online) The atomic and molecular visibilities $\mathcal V_a$ and $\mathcal V_m$. Because of the divergence of $n_a(0)$ or $n_m(0)$, $\mathcal V_a$ (and/or $\mathcal V_m$) must converge to
      unity when an atomic (and/or a molecular) condensate occurs.
    }
  \label{Visibility}
\end{figure}

While it is hard to give a phase diagram with precise borders delimiting the different phases (because each point of the diagram requires a heavy finite-size
scaling analysis), we give a qualitative diagram in Fig.\ref{Phase-diagram} based on simulations for considerably more values of the detuning $D$ and the interaction
$U$ than in Fig.\ref{SM-Phase} to \ref{AC-Phase}. In addition, we provide connection with future experiments by showing
on Fig.\ref{Visibility} how the atomic and molecular visibilities $\mathcal V_a$ and $\mathcal V_m$ \cite{Bloch} behave when entering the AC phase from the SM phase along
the vertical line $D=-2$. In the present case $\mathcal V_a$ reaches unity before $\mathcal V_m$ as the interactions decrease, showing again that the system
is entering an AC phase.

To conclude, we have studied a two-species Bose-Hubbard Hamiltonian for atoms and molecules on a lattice, interacting via a Feshbach
resonance. We have
shown that the MI phase identified in MF studies is actually a SM phase. For large negative detuning we find a transition from this insulating phase to a MC phase, in
agreement with MF theory. For smaller negative or positive detuning however, while MF theory predicts a direct transition from MI to AC+MC, we
find that an AC phase occurs and that the system undergoes phase transitions from SM to AC to AC+MC. The AC phase was not found in previous studies.
The phase diagram we provide may serve as a guide for the investigation of atomic and molecular quantum matter now that condensation of Feshbach molecules is
begining to be achieved experimentally.

\begin{acknowledgments}
We would like to thank Fr\'ed\'eric H\'ebert for useful conversations.
This work is part of the research program of the 'Stichting voor Fundamenteel Onderzoek der Materie (FOM)', which is financially supported
by the 'Nederlandse Organisatie voor Wetenschappelijk Onderzoek (NWO)'.
\end{acknowledgments}

\end{document}